\documentclass[aps,pra,twocolumn,groupedaddress,superscriptaddress,longbibliography]{revtex4-1}
\usepackage{graphicx}
\usepackage{dcolumn}
\usepackage[dvipsnames]{xcolor}
\usepackage{amsmath}
\usepackage{physics}
\usepackage{amssymb}
\usepackage{appendix}
\usepackage{hyperref}
\newcommand{\mycomment}[1]{}
\usepackage{hhline}
\usepackage{longtable}

\begin{document}

\title{Anomalous pinch in electron-electron beam collision}

\author{W. Zhang}
\email[]{wenlong.zhang@ecut.edu.cn}
\affiliation{Jiangxi Province Key Laboratory of Nuclear Physics and Technology, East China University of Technology, Nanchang 330013, China}
\affiliation{Engineering Research Center of Nuclear Technology Application, Ministry of Education, East China University of Technology, Nanchang 330013, China}

\author{T. Grismayer}
\email[]{thomas.grismayer@tecnico.ulisboa.pt}

\author{L. O. Silva}
\email[]{luis.silva@tecnico.ulisboa.pt}
\affiliation{GoLP/Instituto de Plasmas e Fusão Nuclear, Instituto Superior Técnico, Universidade de Lisboa, Lisboa, Portugal}

\date{\today}
\begin{abstract}
{We show that an anomalous pinch can occur in ultra-relativistic electron-electron or positron-positron beam interaction, caused by the combined interplay of collective beam motion (disruption) and strong-field quantum electrodynamics (SF-QED). The locally created electron-positron pairs, from SF-QED effects, screen the self-fields of the beams and can invert the polarity of the Lorentz force, resulting in a pinch of the beams. A theoretical model predicts the pinch condition and is confirmed by first-principles 3-dimensional particle-in-cell simulations. This anomalous pinch causes density compression, increases the collision luminosity, and amplifies the local magnetic fields and the quantum parameter of the beam particles by several orders of magnitude.}
\end{abstract}


\maketitle

\section{Introduction}
\label{sec: introduction}

Lepton accelerators are indispensable tools for many disciplines, including the frontier particle physics \cite{ellis2018future, Shiltsev2021, Gray2021, EurStrParPhys2022}. Future linear lepton colliders \cite{Schroeder2012, ellis2018future, Shiltsev2021, Gray2021, EurStrParPhys2022, Geddes2022, Roser2023, Schroeder2023, Shiltsev2024, ALEGRO2024, Foster2025} will access and probe new physics regimes \cite{Shiltsev2021, Gray2021, EurStrParPhys2022, P5_Report, ILC_Snowmass21, CLIC_Physics}, resorting to unprecedented beam parameters \cite{Schroeder2012, Schulte2017, Shiltsev2021, Geddes2022, Roser2023, Barklow2023, Schroeder2023}. These conditions also open the way to novel platforms for studying the strong-field quantum electrodynamics (SF-QED) \cite{Fabrizio2019, Yakimenko2019, Zhang2023, Tamburini2021}, including the non-perturbative regime \cite{Yakimenko2019}. 
For these future colliders, the collective processes of the beams at the Interaction Point (IP) \cite{Chen1988, Katsouleas1990}, can deteriorate the beams and produce secondary particles that might hinder the outcome of the experiments; the ultra-relativistic particles can be subject to intense electromagnetic fields ($E/E_s > 10^{-4}$ with $E_s$ the Schwinger field) of the oncoming beam, leading to beamstrahlung \cite{Chen1992, Barklow2023, Zhang2023} and $e^-e^{+}$ pair creation \cite{Chen1989, Zhang2023}. In addition, the beams can also be subject to disruption, i.e., the collective transverse motion of the beams \cite{Chen1988}, quantified by the parameter
\begin{equation}
    D = \eta\frac{N_0r_e\sigma_z}{\gamma \sigma_0^2}= \frac{3}{2}\eta\frac{ N_0[10^{10}]\ \sigma_z[\mu m]}{\mathcal{E}_0[\mathrm{GeV}] \ (\sigma_0[0.1\mu m])^2 },
    \label{Eq: D_definition}
\end{equation}
where $N_0$ is the number of particles, $r_e=e^2/mc^2$ the classical electron radius (with the usual definition of the electron charge, mass, and the speed of light), $\mathcal{E}_0=\gamma mc^2$ with $\gamma$ the Lorentz factor, $\sigma_0$ and $\sigma_z$ the transverse width and longitudinal length, respectively. Disruption becomes dynamically significant when $D \geq 1$ with $\eta$ depending on the beam profile: $4$ for a uniform beam and $1$ for a Gaussian beam \cite{Fabrizio2019}. Disruption leads to oscillations of the beams around the propagation axis in collisions of opposite-charge ($e^-e^{+}$) beams, and defocusing of beams in identical-charge ($e^-e^{-}$ or $e^+e^{+}$) collisions. Previous studies in lepton collisions, and lepton collider designs, considered the limits of either low disruption \cite{Noble1987, Chen1989, Chen1992, Raimondi1995} or weak SF-QED effects (beamstrahlung and $e^-e^{+}$ creation) \cite{Chen1988, Barklow1999, ILC_TDR, ILC_Snowmass21, CEPC_beambeameffects}, where the disruption and SF-QED are assumed to be independent processes and can therefore be dealt with separately. This assumption breaks down for future colliders, where the colliding beams possess $\mu m$-scale length and $n m$-scale width \cite{Schulte2017, Shiltsev2021, Schroeder2010, Roser2023, Schroeder2023, Barklow2023, 10TeV_design_initiative} with $D \gg 1$ and the quantum parameter $\chi=\sqrt{(\gamma {\bf {E}}+ {\bf {p}}/mc\times {\bf {B}})^2-( {\bf {p}}/mc\cdot {\bf {E}})^2}/E_s$ larger than unity \cite{Zhang2023} where $\bf {E,\ B}$ are the electromagnetic fields, $E_s=m^2c^3/e\hbar$, $\hbar$ is the reduced Planck constant, and ${\bf {p}}$ is the particle's momentum. The disruption will affect the beam profiles and fields: it was shown that even a mild disruption ($D \sim 1$) in $e^-e^{-}$ collisions would reduce the beamstrahlung as compared to $D \ll 1$ \cite{Zhang2023}. This is due to the dilution of the beams, which causes a decrease in the fields. Previous studies also showed that SF-QED processes can affect the disruption dynamics \cite{ZhangEPS2021, Samsonov2021, Zhang2023}. However, the effect of copious pair production expected when $\chi \gg 1$ and $D \gg 1$ has not been investigated. 

As we show in this paper, the self-consistent description of these processes for future collider parameters leads to novel collective dynamics: the pairs screen the self-fields of the beams and can invert the polarity of the Lorentz force, and this force inversion leads to the anomalous pinch (AP) of the beams, even in the collision of identical-charge beams ($e^- e^-$ or $e^+ e^{+}$), enhancing the luminosity. This anomalous pinch belongs to a broad class of scenarios that share similar underpinning physics, and where the density of produced pairs becomes high enough to modify the background fields, which in turn can modify and even quench QED effects. For instance, in laser-electron scattering, it leads to frequency up-shifts \cite{Zhang2021,Qu2021}, and in laser-driven QED cascades, to significant laser absorption \cite{Grismayer2016, Nerush}, whereas, in neutron-star-associated cascades, it triggers plasma waves and the emission of radio waves and gamma rays \cite{philippov_2020, timokhin_2010, cruz_2021c, Beloborodov2013}.

\section{Theoretical model}
\label{sec: model}
\begin{figure}
\includegraphics[width=7.0cm,height=3.95cm]{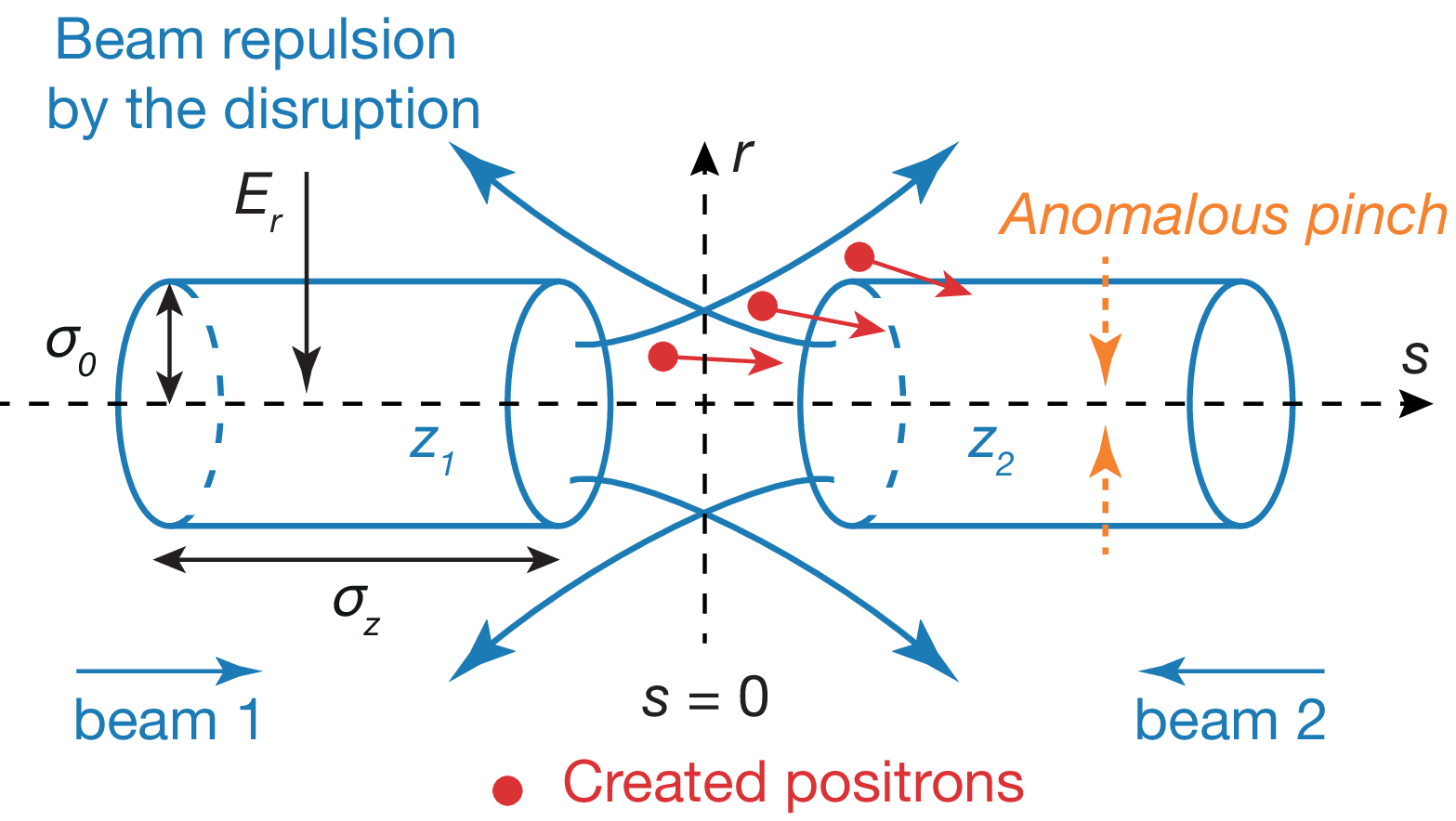}
\caption{(Color online). Schematic of the collision. The colliding (electron) beams are labeled as beam 1 and beam 2. The coordinate $s$ \cite{Chen1988} represents the longitudinal coordinates fixed in the center-of-mass frame. $z_j$ ($j=1,2$) denotes the longitudinal position co-moving with the two beams. The slice at $z_1$ of beam 1 will meet the slice at $z_2$ of beam 2, at the location $s$ and time $t$ which are bound by $s=z_1+ct=-z_2-ct$.}
\label{fig: schematic}
\end{figure}

We consider $e^-e^-$ collisions. A sketch of the collision is represented in Fig. \ref{fig: schematic}, where the beams are assumed to be cold and cylindrical, with uniform density $n_0$ with initial profiles $n_j = n_0$ ($j=1,2$) for $r\leq \sigma_0$, $-\sigma_z\leq z_j\leq 0$. For standard beam parameters from lepton accelerators, $N_0 \sim 10^{10}$, $\sigma_z \gtrsim 10^{-6}~\mathrm{m}$, and $100~\mathrm{GeV} < \mathcal{E}_0 < 10~\mathrm{TeV}$, the SF-QED regime $\chi \gg 1$ is accessible when $\sigma_0 \ll \sigma_z$ \cite{Zhang2023}. Remarkably, this set of parameters also implies high disruption, as this can be verified using Eq. \eqref{Eq: D_definition} when $\sigma_0 < 10~\mathrm {nm}$.

The regime $D\gg1 $ and $\chi \gg 1$, with the aforementioned beam parameters, leads to the time scale hierarchy: $\tau_{col} \gg \tau_\textrm{\tiny D}$, and $\tau_{col} \gg\tau_{\textrm{\tiny QED}}$, where $\tau_{col}=\sigma_z/c$ is the collision time, $\tau_\textrm{\tiny D}= D^{-1/2}\tau_{col}$ is the characteristic disruption time, and $\tau_{\textrm{\tiny QED}}$ ($\propto \chi^{-2/3}$) is the typical photon emission or pair production time (which are on the same order when $\chi \gg 1$) \cite{Ritus1985, Piazza2012, Gonoskov2022}. Depending on the beam parameters, the collision can be disruption-dominated $\tau_\textrm{\tiny D} < \tau_{\textrm{\tiny QED}}$ or QED-dominated $\tau_\textrm{\tiny D} > \tau_{\textrm{\tiny QED}}$. This collision regime presents the benefit of having the response of the newly created pairs within the full interaction time, which can thus affect the self-fields of the beams. In our model, the two time scales are assumed to be on the same order. It permits accounting for beam dilution and pair creation in a regime where these two phenomena are weakly coupled. When the pairs are created before the beams have been significantly disrupted, the dilution of beams and the density of pairs can be computed with the undisrupted beam self-fields given by  
\begin{subequations}
\begin{align}
   E_{r,0} &= -2\pi e n_0r \ \ \ \ \ \ \ \mathrm{for}~~ r\le \sigma_0~(\mathrm{beam}) ; \label{Eq: Er_beam} \\
   E_{r,0} &= -2\pi e n_0\frac{\sigma_{0}^2}{r} \ \ \ \ \ \mathrm{for}~~ r > \sigma_0~(\mathrm{vacuum}). \label{Eq: Er_vacuum}
\end{align}
\label{Eq: Er_expression}
\end{subequations}
The strength of the azimuthal magnetic field satisfies $|B_{\theta,0}| \simeq |E_{r,0}|$ when $\gamma \gg 1$.

\subsection{Beam dilution driven by the disruption}
\label{subsec: beam_dilution}
When beamstrahlung is discarded, the particle energy is constant during collision. Under the undisrupted fields, the equations of motion are
\begin{subequations}
\begin{align}
    \frac{d^2r}{dt^2}&=\frac{\omega_{b}^2}{\gamma } r ~~~~~~~\mathrm{for}~~~r \le \sigma_0~(\mathrm{beam}) ; \label{Eq: equation_motion_beam}\\
    \frac{d^2r}{dt^2}&=\frac{\omega_{b}^2}{\gamma }\frac{\sigma_0^2}{r} ~~~~~\mathrm{for}~~~r > \sigma_0~(\mathrm{vacuum}), \label{Eq: equation_motion_vacuum}
\end{align}
\label{Eq: equation_motion}
\end{subequations}
where $\omega_b=\sqrt{4\pi e^2n_0/m}$ is the beam plasma frequency.
The trajectory of an electron that remains within the beam, at $z_j$ with the initial position $r_{0,j}$, is given by $r_j(z_j,t)=r_{0,j}\cosh(\Delta t_j/\tau_\textrm{\tiny D})$. Here, $\tau_\textrm{\tiny D}=\sqrt{\gamma}/\omega_b$, $\Delta t_j=t-t_{0,j}$, and $t_{0,j}=-z_j/2c$ is the time when the electron crosses the front of the other beam. This trajectory solution indicates that particles are blown away, as sketched in Fig. \ref{fig: schematic}, and that a beam slice expands at a time scale $\tau_\textrm{\tiny D}$. When $t <  \tau_\textrm{\tiny D}$, we can expand the trajectory solution and obtain $r_j(z_j,t)\simeq r_{0,j} [1+\frac{1}{2}D(\Delta t_j/\tau_{col})^2]$. With the equation of continuity $n dr^2=n_0 dr_0^2$, the beam densities read
\begin{equation}
n_{1,2}(s,t) = \frac{n_0} {\left[1+\frac{1}{2}D\left(\Delta t_{1,2}/\tau_{col}\right)^2\right]^2},
 \label{Eq: diluted_density}
\end{equation}
with $\Delta t_{1,2} =\frac{1}{2}(t\pm\frac{s}{c})$, $ct-\sigma_z \le s\le ct$ for beam 1, and $-ct \le s\le \sigma_z-ct$ for beam 2. 

\subsection{Electron-positron pair production}
\label{subsec: pair_creation}

Photon emission and pair production in the SF-QED regime \cite{Ritus1985, Piazza2012, Gonoskov2022, Fedotov2022} are characterized by the quantum parameter $\chi$.  In beam-beam collisions, the electron quantum parameter is $\chi_e(r)=(r/\sigma_0)\chi_{e \, \mathrm{max}}$, where $\chi_{e \, \mathrm{max}}=4\pi e\gamma n_0\sigma_0/E_s$. When the two beams start colliding, pair production follows an SF-QED shower behavior because the particle trajectories remain almost perpendicular to the self-fields. Before the onset of disruption, and for $\chi$ up to a few 10s, only the first generation of pairs is relevant \cite{Pouyez2024,pouyez2024multiplicity}. If the radiation cooling of the electron beam is discarded, the density of pairs around the beam front is given by \cite{Pouyez2024}
\begin{equation}
n_p(r,t) = n_0\int_0^tdt' \int_0^{\chi_e(r)} d\chi_\gamma \frac{d^2W}{dtd\chi_\gamma} \left[1-e^{-W_p(\chi_\gamma)(t-t')}\right] ,
\label{density1}
\end{equation}
where $d^2W/dtd\chi_\gamma$ is the differential probability rate of photon emission \cite{Zhang2023}, $\chi_\gamma = \xi \chi_e$ is the quantum parameter for photons, and $\xi=\mathcal{E}_\gamma/\mathcal{E}_0$ the normalized photon energy. $W_p (\chi_\gamma)$ is the rate of pair creation from a photon with $\chi_\gamma$ \cite{Ritus1985}. Expanding the exponential in Eq. \eqref{density1} at the first order, the density reads
\begin{equation}
 n_p(r,t) \simeq \frac{1}{2}n_0\mathcal{R}^2(r) t^2 ,  
 \label{density}
\end{equation}
where $\mathcal{R}^2(r)= \int_0^{\chi_e(r)} d\chi_\gamma \left( d^2W/dtd\chi_\gamma \right) W_p(\chi_\gamma)$, and $\mathcal{R}$ represents the pair rate at each $r$, averaged over the photon spectrum. In the limit $\chi_e \gg 1$, $\mathcal{R}(r)\simeq \sqrt{\ln{\chi_e(r)}}\chi_e^{-1/3}(r)W_\gamma (r)$, where $W_\gamma(r)=\int_0^{\chi_e (r)} d\chi_\gamma d^2W/dtd\chi_\gamma  \simeq 1.46\alpha /(\tau_c\gamma) \chi_e^{2/3}(r)$ \cite{Ritus1985} with $\tau_c = \hbar/mc^2$ the Compton time and $\alpha$ the fine-structure constant. Equation \eqref{density} is an excellent approximation of the exact pair density until $t \simeq 5W_\gamma^{-1}$ \cite{Pouyez2024} and is therefore suitable for this model.

\subsection{Anomalous pinch}
\label{subsec: AP_condition}

The new electrons are expelled from the initial beam region, whereas the positrons remain confined (the opposite will occur in a $e^{+}e^{+}$ collision). The effect of the increasing positron population is to screen the self-fields of the beams and then invert the polarity of the Lorentz force. This inversion causes a pinch of the beam electrons, more pronounced in the middle and tail parts, as sketched in Fig. \ref{fig: schematic}. This anomalous pinch (AP) effect has been reported in CAIN simulations for the design of a Collider Higgs Factory \cite{barklow2023xcc}. The pinch onset can be quantified using our previous results. At the front of beam $1$ (with $s=ct$), the expelling force exerted on beam $2$ is $F(r,t)=4\pi e^2n_e(t)r$, where $n_e(t)=n_0[1+D(t/\tau_{col})^2/2]^{-2}$ from Eq. \eqref{Eq: diluted_density}. The charge separation of new electrons and positrons is first assumed here to occur almost instantaneously. In that case, the focusing force, from the positrons and experienced by beam $2$, is $F_p(r,t)=-2eE_p(r,t)$, where $E_p(r,t) = 4\pi e\int_{r_0}^rn_p(r',t)r'dr'/r$ and $r_0=\sigma_0/\chi_{e \, \mathrm{max}}$ (where the local $\chi_e$ is unity). The focusing force at $r = \sigma_0$ is given by  
\begin{equation}
  F_p(\sigma_0,t)\simeq -\frac{3\pi}{2} e^2 n_0\sigma_0\mathcal{R}^2(\sigma_0)t^2 .
\end{equation}
The pinch can occur if the polarity of the total force is inverted, i.e., $F_{tot}=F(\sigma_0,t)+F_p(\sigma_0,t)< 0$. The time $t_F$ corresponding to $F_{tot}= 0$ is obtained by solving 
\begin{equation}
    \frac{3}{8}\mathcal{R}^2(\sigma_0)t_{F}^2 \left(1+\frac{1}{2}D \frac{t_{F}^2}{\tau_{col}^2} \right)^2=1 ,
\end{equation}
whose solution is
\begin{equation}
    \frac{t_{F}}{\tau_{col}} = \frac{1}{\sqrt{2D}}\left(\sqrt{1+\frac{32}{3}\frac{D}{\mathcal{R}^2(\sigma_0)\tau_{col}^2}} - 1 \right)^{1/2} .
    \label{Eq: t_F}
\end{equation}
At $t_F$, the positron density is 
\begin{equation}
    \frac{n_p}{n_0} = \frac{4}{3}\left(1+\frac{1}{2}\left(\frac{t_F}{\tau_\textrm{{\tiny D}}}\right)^2\right)^{-2} .
    \label{den_pos}
\end{equation}
Since $t_F< \tau_\textrm{{\tiny D}}$ by the validity of Eq. \eqref{Eq: diluted_density}, Eq. \eqref{den_pos} shows that the density of positrons needed to screen the field is on the order of the initial beam density. The solution for $t_F$ was derived under the assumption of instantaneous charge separation of the pairs, which amounts to considering $\tau_\textrm{{\tiny D}} \lesssim \tau_{\textrm{\tiny QED}}$. For a fixed $\chi_{e \, \mathrm{max}}$, Eq. \eqref{Eq: t_F} should be all the more a good estimate of the AP onset for high values of $D$, since one has
\begin{equation}
    \frac{32}{3}\frac{D}{\mathcal{R}^2(\sigma_0)\tau_{col}^2} \sim \frac{32}{3}\left(\frac{\tau_{\textrm{\tiny QED}}}{\tau_\textrm{{\tiny D}}}\right)^2\gg 1 .
\end{equation}
In this limit, we obtain
\begin{equation}
 t_{F} \sim \sqrt{\tau_\textrm{{\tiny D}}\tau_{\textrm{\tiny QED}}} .
 \end{equation}
In the realistic case, the charge separation of the created electrons and positrons occurs in the typical disruption time $\tau_\textrm{\tiny D}$, as observed in mild-disruption collisions ($ D\sim1$) in a previous study \cite{Zhang2023}. The onset of the pinch (characterized by $t_{\textrm{\tiny AP}}$) should be bounded as $t_F<t_{\textrm{\tiny AP}}\lesssim t_F+\tau_\textrm{\tiny D}$. Our numerical study shows that notable pinching is observed when $t_{\textrm{\tiny AP}}\lesssim \tau_{col}/2$ which implies $t_F+\tau_\textrm{\tiny D} \le \tau_{col}/2$. For $D\gg 1$, this criterion for pinch formation can be recast as $\mathcal{E}_0[\mathrm{GeV}]/\sigma_z[\mu m] \le 5D^{1/4}\chi_{e \, \mathrm{max}}^{1/3}$ which can be further translated into a convenient form as
\begin{equation}
    \frac{(\mathcal{E}_0[\mathrm{GeV}])^{11/12}\ (\sigma_0[0.1\mu m])^{5/6}}{(\sigma_z[\mu m])^{11/12}(N_0[10^{10}])^{7/12}}\le 7.
    \label{Eq: AP_condition_engineering}
\end{equation}

Although the theoretical model has been established for uniform and cylinder-shaped beams, the results can be conveniently applied to realistic Gaussian-profile beams. The profile transform was proposed in a recent publication \cite{Zhang2023}. Using the  conservation of particle flux and the quantum parameter $\chi$ between the Gaussian and uniform beams, we obtain the profile transform as 
\begin{equation}
    \sigma_{z}^{\mathrm{U}}=2\sqrt{2}\sigma_{z}^{\mathrm{G}}, \ n_{0}^{\mathrm{U}}=0.41n_{0}^{\mathrm{G}}, \ \sigma_{0}^{\mathrm{U}}=2.22\sigma_{0}^{\mathrm{G}},
    \label{Eq: profile_transform}
\end{equation}
where the superscript ``G" represents the Gaussian beam, and ``U" the equivalent uniform beam. One can use this profile transform for Eqs. \eqref{Eq: t_F}, \eqref{Eq: AP_condition_engineering}, and \eqref{Eq: theoretical_HD}. The detailed analysis is provided in Appendix \ref{SM_subsec: profile_transform}.

\subsection{Other regimes of interaction}
\label{subsec: other_regimes}

We have considered for the derivation of the AP, the special time scale hierarchy $\tau_\textrm{\tiny D} \sim \tau_{\textrm{\tiny QED}} \ll \tau_{col}$. We could also ask whether a pinch would occur when these two characteristic times are not on the same order, keeping in mind that $\chi \gg 1$, and $ D\gg1$.

The case $\tau_{\textrm{\tiny QED}} \ll \tau_\textrm{\tiny D}$ implies that the beam energy cannot be too high, typically $\mathcal{E}_0 < 100~\mathrm{GeV}$, ruling out the possibility of observing the effect on a modern collider, but wakefield-accelerated beams may apply. The other option is to have the quantum parameter arbitrarily high, which poses two crucial problems. SF-QED is fully non-perturbative, with no existing theory above $\alpha^{2/3}\chi > 1$ \cite{Yakimenko2019}. The extreme values of $\chi$ with this beam energy range require an arbitrarily small transverse beam size ($\sigma_0 < \mathrm{nm}$), and such beams may not be conceivable with current technology. From a physics perspective, it also leads to the strange scenario where many pairs are created before being separated by the self-fields. If the multiplicity of the showers turned out to be large, $n_p/n_0 \sim \chi_e$ \cite{Pouyez2024}, the new pairs would completely dominate the collision, and there would be no reason to observe a pinch.

The opposite case $\tau_\textrm{\tiny D} \ll \tau_{\textrm{\tiny QED}}$ supposes a significant dilution of the beams before new pairs are created. Due to the decrease in the density (and the field associated), the density of positrons needed to invert the Lorentz force should be low. This case is harder to solve analytically, but we can still estimate the onset of the pinch. The density of the beam front evolves now as $n = n_0/\cosh^{2}(t/\tau_\textrm{\tiny D})\simeq n_0 e^{-2t/\tau_\textrm{\tiny D}}$, and the quantum parameter in Eq. (\ref{density1}) as $\chi_e(t) = \chi_e(0)/\cosh(t/\tau_\textrm{\tiny D})\simeq \chi_e(0)e^{-t/\tau_\textrm{\tiny D}}$. Noticing that the function $\mathcal{R}$ depends weakly on $\chi$, namely $\mathcal{R}(t) \sim W_{\gamma}(t)$, the positron density reads for $t \gg \tau_\textrm{\tiny D}$
\begin{eqnarray}
    \frac{n_p}{n_0} & = & \int_0^t dt' \mathcal{R}^2(t')(t-t') \\
    & \sim & W^2_{\gamma}(0) t \tau_\textrm{\tiny D} \sim \frac{t \tau_\textrm{\tiny D}}{\tau^2_{\textrm{\tiny QED}}} .
\end{eqnarray}
Following the method of Sec. \ref{subsec: AP_condition}, the time $t_F$ is found by solving the equation
\begin{equation}
    W^2_{\gamma}(0) t_F \tau_\textrm{\tiny D}e^{2t_F/\tau_\textrm{\tiny D}}\sim 1 .
\end{equation}
The solution is
\begin{equation}
    t_F \sim \frac{\tau_\textrm{\tiny D}}{2}W\left(\frac{{\tau^2_{\textrm{\tiny QED}}}}{2\tau^2_\textrm{\tiny D}}\right)\rightarrow \frac{\tau_\textrm{\tiny D}}{2}\ln\left(\frac{{\tau^2_{\textrm{\tiny QED}}}}{2\tau^2_\textrm{\tiny D}}\right) ,
\end{equation}
where $W$ is the Lambert function, known as the product algorithm. Even for a large separation time scale between $\tau_\textrm{\tiny D}$ and $\tau_{\textrm{\tiny QED}}$, $t_F$ remains a few $\tau_\textrm{\tiny D}$. The corresponding positron density is 
\begin{equation}
\frac{n_p}{n_0} \sim \frac{\tau^2_\textrm{\tiny D}}{\tau^2_{\textrm{\tiny QED}}}\ln\left(\frac{{\tau^2_{\textrm{\tiny QED}}}}{2\tau^2_\textrm{\tiny D}}\right) \ll 1 .
\end{equation}
It is thus possible to observe a local inversion of the Lorentz force in this regime. Nonetheless, the associated pinch will be almost insignificant due to the very few positrons involved in the process.

\begin{figure}
\includegraphics[width=6.5cm,height=5.93cm]{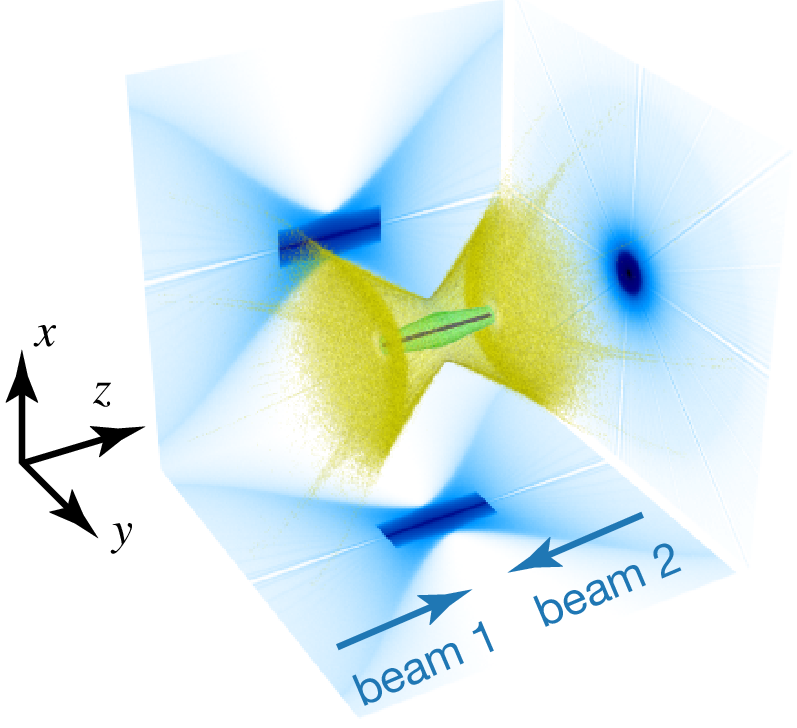}
\caption{(Color online). 
 Anomalous pinch in a 3D PIC simulation of an electron-electron collision. The colors represent different iso-density contours. The yellow region ($n/n_0 = 0.1$) shows the expanding fronts of the beams. The green ($n/n_0 = 1$) and purple ($n/n_0 = 10$) regions show the core layers of the pinched beams. The beam parameters are: $\mathcal{E}_0=70 \ \mathrm{GeV}$, $\sigma_z=12\ \mathrm{\mu m}$, $\sigma_0=4.9\ \mathrm{nm}$, and $N_0=1.3\times 10^{9}$, corresponding to $D=54$ and $\chi_{e \, \mathrm{max}}=13$.}
\label{fig: 3D_density}
\end{figure}

\section{Three-dimensional (3D) particle-in-cell (PIC) simulations}
\label{sec: PIC_simulations}

The AP effect proposed above has been investigated by full-scale 3D PIC simulations using \small{OSIRIS} code \cite{OSIRIS}, where the SF-QED processes are self-consistently included. The QED-PIC framework with {\small{OSIRIS}} is described in Appendix \ref{SM_sec: QED_OSIRIS}. The simulation result is illustrated in Fig. \ref{fig: 3D_density}. The colliding beams are cylinder-shaped with uniform density as utilized in our model (Sec. \ref{sec: model}). The beams are pinched due to the AP effect, which is demonstrated by the core layers (green and purple regions in Fig. \ref{fig: 3D_density}). The simulation box $(x,y,z)$ is $125\ \mathrm{nm}\times 125\ \mathrm{nm}\times 24\ \mathrm{\mu m}$ resolved by $300\times 300\times 30000$ grids, leading to the numerical resolution of $\Delta x=\Delta y = 0.08\sigma_0$ and $\Delta z=6.7\times 10^{-5}\sigma_z$. $4$ particles-per-cell (PPC) are used, corresponding to $2.6\times 10^{7}$ macro-particles per beam. The time step is $\Delta t=2.2\times 10^{-5}\sigma_z/c$. The locally constant field approximation (LCFA) well holds for the collision studied here (the condition for valid LCFA in a quantum-dominated beam-beam collision is given in Ref. \cite{Zhang2023}).

\begin{figure}
\includegraphics[width=6.5cm,height=5.4cm]{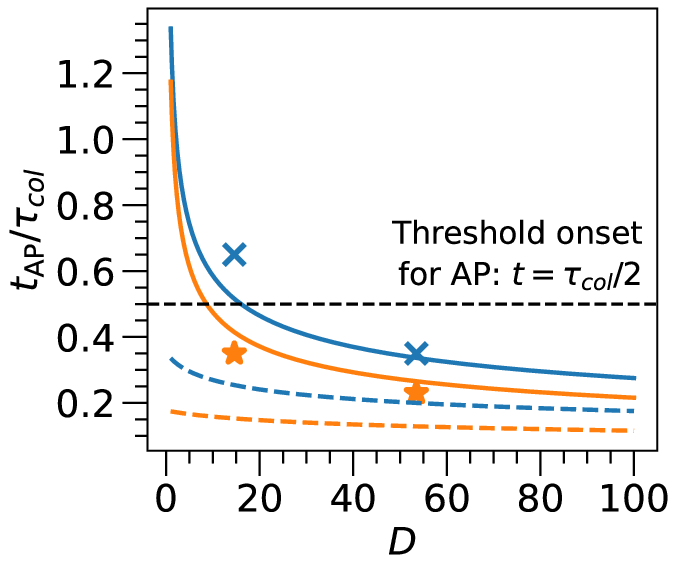}
\caption{(Color online). Onset time of AP ($t_{\textrm{\tiny AP}}$) as a function of the disruption parameter for different beam lengths, measured from PIC simulations (symbols) and compared with theoretical predictions and the typical threshold formation time ($t=\tau_{col}/2$). The simulations are performed for $D=14.6$ and $54$, respectively. The blue and orange colors are for $\sigma_z=6\ \mathrm{\mu m}$ and $12\ \mathrm{\mu m}$, respectively. We maintain $\mathcal{E}_0=70 \ \mathrm{GeV}$, and choose the density $n_0$ and waist $\sigma_0$ to keep $\chi_{e \, \mathrm{max}}=13$ constant. The dashed lines depict $t_F/\tau_{col}$ [Eq. \eqref{Eq: t_F}], and solid lines represent $(t_F+\tau_\textrm{\tiny D})/\tau_{col}$. The simulations verify that significant AP occurs at the early stage of collisions (with $t_{\textrm{\tiny AP}}<\tau_{col}/2$), and $t_{\textrm{\tiny AP}}$ lays between $t_F$ and $t_F+\tau_\textrm{\tiny D}$ in agreement with our model.}
\label{fig: AP_condition}
\end{figure}

The theoretical criterion for the AP formation [Eqs. \eqref{Eq: t_F} and \eqref{Eq: AP_condition_engineering}] has been verified by PIC simulations where we have measured the times $t_{\textrm{\tiny AP}}$ when the total transverse force vanishes in the different simulations (Fig. \ref{fig: AP_condition}). For $\sigma_z = 6\ \mu m$, $t_{\textrm{\tiny AP}}$ seems slightly above the curve $t_F+\tau_\textrm{\tiny D}$, whereas for $\sigma_z = 12\ \mu m$, $t_{\textrm{\tiny AP}}$ confirms the analytical results and lays between $t_F$ and $t_F+\tau_\textrm{\tiny D}$. Even if the total force has vanished or switched polarity a significant electron pinch is only observed for the points under $\tau_{col}/2$. This shows that high disruptions (large $D$) with elongated beams (large $\sigma_z$) are required for driving the pinch at an early stage of the collision, such that the pinch has enough time to develop to become noticeable. When both $D$ and $\sigma_z$ are chosen such that $t_{\textrm{\tiny AP}}>\tau_{col}/2$, the AP is too weak to impact the collision dynamics.

We have also compared {\small{OSIRIS}} results with {\small{GUINEA-PIG}} (a specialized beam-beam code \cite{Schulte1996}) for Gaussian beam profiles and obtained excellent agreement without SF-QED and reasonable agreement with SF-QED. An example is shown in Appendix \ref{SM_subsec: OS_GP}, where simulations with both codes give similar beam dynamics and close luminosity enhancement due to the AP phenomenon. The deflection of the beams after they are severely pinched indicates the development of hosing/kink-like instabilities. However, the detailed values of photon emission, pair production, and density compression are slightly different between these two simulations. A thorough and systematic comparison between these two codes is beyond the scope of this paper and will be delivered in future work. To further complement our study, we have also performed simulations (with {\small{OSIRIS}} and {\small{GUINEA-PIG}}) for flat beams with asymmetric transverse profiles, showing that the pinch condition (Eq. \eqref{Eq: AP_condition_engineering}) still holds. We finally stress that realistic finite beam emittance does not significantly change the AP physics shown here, as confirmed by QED-PIC simulations (see Appendix \ref{SM_subsec: impact_emittance}).

\section{Impact of AP on the luminosity}
\label{sec: luminosity}

This pinch also has consequences on the main beam parameters of the IP. The collision luminosity is \cite{Wiedemann2007}
\begin{equation}
L_0= 2c\int dx\int dy\int dz \int dt~n_1 n_2 . 
\end{equation}
It represents the total number of scattering events at the IP. When disruptions are not considered, $L_0$ is the geometric luminosity ($L_0 = L_0^{Geo}$); for cylinder-shaped and uniform beams, $L_0^{Geo} = N_0^2/\pi \sigma_{0}^2=\pi n_0^2 \sigma_{0}^2\sigma_z ^2$. For significant disruption and neglecting the SF-QED, the luminosity of $e^- e^-$ collisions is reduced by the factor $H_D=L_0/L_0^{Geo}$ that can be calculated for mild and large disruptions. There are two ways of looking at the expansion for $t < \tau_\textrm{\tiny D}$ that leads to Eq. \eqref{Eq: diluted_density}. The expansion is valid for short times for all $D$ or during the whole collision time but for mild $D$. With a cylindrical dilution, the expression of the luminosity reduces to $L_0 = 4L_0^{Geo}\int dt \int ds \int n(s,t)/n_0$. For mild disruptions, the dilution is given by Eq. \eqref{Eq: diluted_density}, which gives $ H_D=8/(8+D)$. When $D\gg 1$ (with $\tau_\textrm{\tiny D}\ll \tau_{col}$), the electrons will move beyond the initial volume of the beams for longer times, and then experience the vacuum fields. The asymptotic particle trajectory for $t\gg \tau_\textrm{\tiny D}$ is $r_j(z_j,t)\simeq r_{0,j} [1+\sqrt{D}/2(\Delta t_j/\tau_{col})]$. We can similarly compute the density dilution for $D \gg 1$ and we obtain $H_D$ as
\begin{equation}
    H_D \simeq \frac{32}{D}\left[\ln(1+\frac{\sqrt{D}}{4})-\frac{\sqrt{D}}{\sqrt{D}+4}\right] .
    \label{Eq: theoretical_HD}
\end{equation}

\begin{figure}
\includegraphics[width=8.5cm,height=9.37cm]{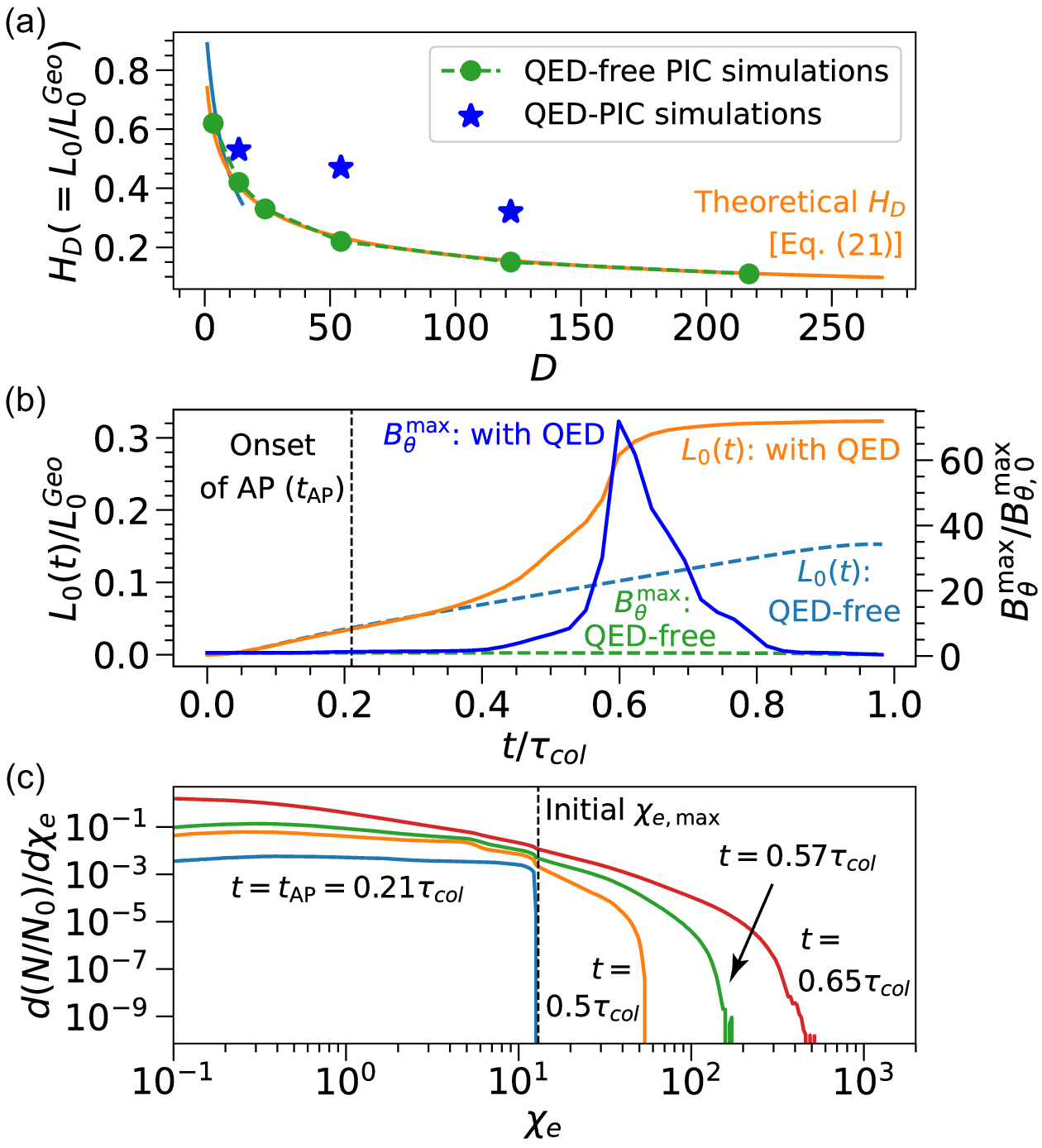}
\caption{(Color online). (a) Luminosity reduction $H_D$ for $e^-e^-$ collisions as a function of the disruption parameter. The solid lines represent the theoretical models. For the mild-$D$ regime (light blue), $H_D= 8/(8+D)$; for the high-$D$ regime (orange), $H_D$ is given by Eq. \eqref{Eq: theoretical_HD}. The 3D simulations with SF-QED switched off show an excellent agreement with the models. These simulations use uniform beams with different $\sigma_z$ while keeping other parameters unchanged, including $\mathcal{E}_0=70 \ \mathrm{GeV}$, $\sigma_0=11.3\ \mathrm{nm}$, and $n_0=6.23\times 10^{23}\ \mathrm{cm}^{-3}$. (b) PIC simulations for an $e^-e^-$ collision with $\sigma_z=27.6\ \mathrm{\mu m}$, $D=122$, and $\chi_{e \, \mathrm{max}}=13$. The other parameters, including $\mathcal{E}_0$, $\sigma_0$, and $n_0$, are the the same with (a). Left axis: the luminosity growth over time ($L_0(t)/L_0^{Geo}$). Right axis: the maximum magnetic field ($B_{\theta}^{\mathrm{max}}/B_{\theta, 0}^{\mathrm{max}}$, with $B_{\theta, 0}^{\mathrm{max}}$ the initial peak field). The measured AP onset ($t_{\textrm{\tiny AP}}=0.21\tau_{col}$) is indicated. (c) Evolution of the $d(N/N_0)/d\chi_e$ distribution in the collision shown in (b). One observes that the distribution develops a tail pushed into the deep-quantum regime by AP.}
\label{fig: Lumi_HD}
\end{figure}

In Fig. \ref{fig: Lumi_HD}(a), the theoretical predictions of $H_D$ are shown to be in excellent agreement with the QED-free PIC simulations. When SF-QED is included (blue stars), the slower decrease in $H_D$ is a manifestation of AP. When $D$ is mild (for $D = 14$) the onset of AP occurs close to $t \simeq \tau_{col}/2$ (see Fig. \ref{fig: AP_condition}) and the beams cannot be pinched efficiently, resulting in a $H_D$ similar to the QED-free result. For $D \gg 10$, $H_D$ becomes significantly enhanced by AP. 

The enhancement of $H_D$ is a direct consequence of the increase of instantaneous luminosity when AP occurs, as shown in Fig. \ref{fig: Lumi_HD}(b). The onset of the pinch is observed around $t \simeq 0.2 \tau_{col}$ in agreement with the theoretical estimate ($t_F + \tau_\textrm{\tiny D}$). The luminosity with QED effects departs from the QED-free $L_0(t)$ shortly after the AP onset and rises until $t \simeq 0.6 \tau_{col}$ (the pinch stage of the tail of the electron beams). Nevertheless, the pinch cannot go indefinitely because beam instabilities mediated by QED effects \cite{ZhangEPS2021, Samsonov2021} will eventually compete with the compression and destroy the beams. These instabilities, such as kink or hose, have a typical growth rate of $\omega_b/\sqrt{\gamma} = 1/\tau_\textrm{\tiny D}$ \cite{yokoya2005beam}. Even without a thorough study, we can conjecture that this limits the compression stage to a few $\tau_\textrm{\tiny D}$. The saturation of the luminosity (flattening of the orange curve) for $t> 0.6\tau_{col}$ conveys the emergence of these beam instabilities, that we have observed in our simulations (one example is shown by Fig. \ref{fig: OS_GP} in Appendix \ref{SM_subsec: OS_GP}). In addition to the exotic beam movements, the beams also gain a significant energy spread due to the beam-beam effects (see Appendix \ref{SM_subsec: OS_GP} for details of the spread). There will be a trade-off between the overall luminosity enhancement and the broadened energy spread which can diminish the luminosity at specific center-of-mass energy designated by the particle colliders.

\section{Discussion and Conclusion}
\label{sec: discussion_conclusion}

We showed that the interplay between high disruption and SF-QED effects in ultra-relativistic $e^- e^-$ collisions results in a surprising modification of the collective beam dynamics. The SF-QED shower of positrons can reach a density capable of screening the self-fields and inverting the initially repelling Lorentz force acting on the electrons, eventually leading to an anomalous pinch of the beam. The pinch can be efficient for sufficiently long ($\sigma_z \gtrsim \textrm{10's}~\mathrm{\mu m}$), and thin ($\sigma_0 \lesssim \textrm{10's}~ \mathrm{nm}$) beams. 

As the beam compresses, the self-fields of the beam also increase, as shown in Fig. \ref{fig: Lumi_HD}(b). This effect of strong self-field amplification and density compression was first described in $e^-e^+$ collisions in \cite{SorboAPS2019}. Using particle conservation, the pinched density scales as $1/\sigma_{0}^2(t)$ with the beam waist $\sigma_0(t)$ decreasing during AP. The associated electromagnetic fields and the quantum parameter are enhanced as $1/\sigma_{0}(t)$. Therefore, the AP can be used to access the frontier of the non-perturbative SF-QED regime \cite{Yakimenko2019} for fundamental studies. As an example, the peak magnetic field is increased by more than $70$ times in the pinching areas of the collision as shown in Fig. \ref{fig: Lumi_HD}(b). For beam electrons that have not suffered radiation loss, their $\chi_e$ are raised by the same magnetic-field-compression factor, showing $\chi_e \gtrsim 500$. This $\chi_e$ amplification is illustrated in Fig. \ref{fig: Lumi_HD}(c), which displays the $\chi_e$ distribution at four times. The distribution, initially peaked, spreads out during the pinch, and allows some electrons to approach the non-perturbative SF-QED regime ($\alpha\chi_e^{2/3} \sim 1$).

The AP discovered here is relevant to the IP of future colliders and must be considered for the collider design. A summary of near-future projects of linear colliders where AP can be observed \cite{Schulte2017, Shiltsev2021, Roser2023, Schroeder2023, Barklow2023, 10TeV_design_initiative} is provided in Appendix \ref{SM_subsec: future_colliders}. The anomalous pinch represents a tangible manifestation of the back-reaction of pair creation on the self-fields of the beams. Whereas the pinch might not be directly observed in an experiment, the outcome of the scattering events determined by the luminosity would be enhanced by the pinch of the beams. This opens a novel route for future colliders based on $e^-$ beams only \cite{ALEGRO2024, Foster2025, 10TeV_design_initiative}.

\begin{acknowledgments}
We want to acknowledge discussions with Dr. D. Del Sorbo, Professor F. Fiuza, Professor W. B. Mori, Dr. Tim Barklow, and Dr. Glen White. This work was supported by FCT (Portugal) Grants No. 2022.02230.PTDC (X-MASER), UIDB/FIS/50010/2020 - PESTB 2020-23, Grants No. CEECIND/04050/2021 and PTDC/FIS-PLA/ 3800/2021. W.Z. is supported by Jiangxi Provincial Natural Science Foundation (the Young Scientists Fund, No. 20242BAB21003), the Start-up Fund (No. DHBK2023009) from East China University of Technology (ECUT, Nanchang, China), and the East China Accelerator $\&$ Neutron Source (ECANS) project at ECUT. We acknowledge EuroHPC for awarding access to Karolina supercomputer in Czech Republic and MareNostrum at Barcelona Supercomputing Center (BSC, Spain). Simulations were performed at National Supercomputer Center in Guangzhou (China), Karolina (Czech Republic), and MareNostrum (Spain).
\end{acknowledgments}

\appendix
\section{QED-PIC simulation framework}
\label{SM_sec: QED_OSIRIS}

\subsection{PIC and SF-QED}
\label{SM_subsec: QED_PIC_overview}
Particle-in-cell (PIC) codes are one of the most important research tools in plasma physics \cite{Dawson1983, Birdsall1991} as they describe the self-consistent microscopic interaction between a collection of charged particles. The standard loop of the PIC method can be summarized as follows. The simulation domain is represented by a discrete spatial grid, in which macro-particles, representing an ensemble of real particles, move continuously. As they move across the grid, charged macro-particles carry electrical currents deposited on the grid vertices. These currents, defined with the vector $\bf J$, are then used to advance the electric and magnetic fields $\bf E$ and $\bf B$ in time via Faraday’s and Ampere's laws. The updated electromagnetic field values, defined on the grid vertices, are then interpolated to the particles’ positions and used to compute the Lorentz force acting on them. 

When the plasma is exposed to intense electromagnetic fields, the particle dynamics can enter the strong-field quantum electrodynamics (SF-QED) regime. The SF-QED regime is characterized by the dimensionless parameter $\chi_e$
\begin{equation}
    \chi _e=\frac{1}{E_s}\sqrt{\left(\gamma {\bf {E}}+ \frac{\bf {p}}{mc} \times {\bf {B}} \right)^2-\left( \frac{\bf {p}}{mc}\cdot {\bf {E}} \right)^2},
    \label{Eq: Suppl_chi_definition}
\end{equation}
where $E_s=m^2c^3/e\hbar=4.41\times 10^{13}\ \mathrm{statV/cm}$ is the Schwinger field with $e$ and $m$ are the charge and mass of an electron, $\hbar$ is the reduced Planck constant, $c$ is the speed of light in vacuum, $\gamma$ and ${\bf {p}}$ are the particle's Lorentz factor and momentum. When $\chi_e\gtrsim 1$, the photon emission is purely quantum and the photons have a non-negligible probability of decaying into an electron-positron pair \cite{Gonoskov2022, Fedotov2022}. The SF-QED regime is reached when the ultra-relativistic dense lepton beams are collided. Modeling these collisions requires being able to describe the SF-QED processes and also the collective beam motion due to the self-fields of the beam. 

It is in this way that PIC simulations enriched with a QED module prove to be the ideal tool to study the beam-beam physics at the Interaction Point (IP) of lepton colliders \cite{Fabrizio2019, Zhang2023, Samsonov2021}. The PIC method intrinsically ensures a self-consistent calculation for the evolution of beam motion and fields, as well as the quantum radiation and $e^-e^+$ pair production. This makes PIC simulations a powerful tool for studying beam-beam collisions, especially in the high-disruption and strong-quantum regimes. 

\subsection{SF-QED processes in {\small{OSIRIS}}}
\label{SM_subsec: SFQED_OSIRIS}
{\small{OSIRIS}} \cite{OSIRIS} has been extended to incorporate several SF-QED processes, including nonlinear Compton scattering (NCS) also known as the quantum-corrected synchrotron radiation, and nonlinear Breit-Wheeler (NBW) for $e^-e^+$ pair production. These two phenomena are the leading and most relevant QED processes \cite{Gonoskov2022, Fedotov2022} in strong fields. 

NCS is a self-consistent model describing the interaction between a lepton with ultra-relativistic energy $\mathcal{E}_0=\gamma mc^2$ and a strong background electromagnetic field. In this interaction, the lepton emits a gamma-ray photon with energy $\mathcal{E}_\gamma=\xi\mathcal{E}_0$. Under the locally constant field approximation (LCFA), the differential probability rate of NCS reads
\begin{equation}
\frac{d^2P_\gamma}{dtd\xi}=\frac{\alpha}{\sqrt{3}\pi \tau_c \gamma }\left[\int_{b}^{\infty}K_{5/3}(q)dq + \frac{\xi^2}{1-\xi}K_{2/3}(b) \right]
\label{Eq: Suppl_W_gamma},
\end{equation}
where $\tau_c = \hbar/mc^2$ is the Compton time. $\alpha$ is the fine-structure constant. $b=2/(3\chi_{e})\xi/(1-\xi)$, $K_{\nu}$ the modified Bessel function of the second kind, and $\chi_e$ is quantum parameter given by Eq. \eqref{Eq: Suppl_chi_definition}. The total rate is given by $dP_\gamma/dt=\int_0^1(d^2P_\gamma/dtd\xi)d\xi$. In the strong-quantum regime ($\chi_e\gg 1$), $dP_\gamma/dt\simeq 1.46\alpha/(\tau_c \gamma)\chi_{e}^{2/3}$. NCS is an intrinsically many-body problem where the external fields are treated as a collection of coherent photons and the particle’s spin needs to be accounted for \cite{Gonoskov2022}. For NCS, $\xi = 1$ is theoretically allowed, corresponding to the radiation event with complete energy transfer from the lepton to emitted photon. However, the probability for this photon emission is technically negligible, since $d^2P_\gamma/dtd\xi \propto b^{-2/3}\exp (-b)$ which exponentially decreases to $0$ when $\xi \rightarrow 1$ (see Ref. \cite{Zhang2023} for detailed analysis). The treatment of NCS is therefore different from the linear Compton scattering. For example, in the special regime of full inverse linear Compton scattering, the maximum energy of the scattered photon is bound by $\xi=1-1/2\gamma$ \cite{Serafini2024}.

While the emitted photons propagate in the electromagnetic field, they can decay into an $e^-e^+$ pair through the NBW process. The differential probability rate of NBW pair production is 
\begin{equation}
    \frac{d^2P_{pp}}{dtd\xi^-} = \frac{\alpha mc^2}{\sqrt{3}\pi \tau_c \mathcal{E}_\gamma }\left[\left(\frac{ \xi^+ }{ \xi^-} + \frac{ \xi^-}{ \xi^+} \right)K_{2/3}(b) + \int_{b}^{\infty}K_{1/3}(q)dq  \right],
    \label{Eq: Suppl_W_pp}
\end{equation}
where
\begin{equation}
\xi^- = \frac{\mathcal{E}^-}{\mathcal{E}_\gamma}, \  \xi^+ = \frac{\mathcal{E}^+}{\mathcal{E}_\gamma}, \ b=\frac{2}{3\chi_\gamma}\frac{1}{\xi^-\xi^+},
\label{Eq: Suppl_notation_for_Wpp}
\end{equation}
with $\mathcal{E}^-$ and $\mathcal{E}^+$ being the energies of the new electron and positron, respectively. The quantum parameter $\chi_\gamma$ is similarly defined as
\begin{equation}
    \chi _\gamma =\frac{1}{E_s}\sqrt{\left(\frac{\mathcal{E}_\gamma}{mc^2} {\bf {E}}+ \frac{\hbar \bf {k}_\gamma}{mc}\times \bf {B}\right)^2-\left( \frac{\hbar \bf {k}_\gamma}{mc}\cdot \bf {E} \right)^2},
    \label{Eq: Suppl_chi_gamma_definition}
\end{equation}
where $\bf {k}_\gamma$ is the wave vector of the photon. The quantum parameter of the produced pairs can be obtained as $\chi_{e}^{\pm}=\xi^{\pm}\chi_\gamma$. Similarly, the overall rate of pair production is given by $dP_{pp}/dt=\int_0^1(d^2P_{pp}/dtd\xi^-)d\xi^-$. For the strong-quantum regime ($\chi_\gamma \gg 1$), $dP_{pp}/dt\simeq 0.38\alpha mc^2/(\tau_c \mathcal{E}_\gamma)\chi_\gamma ^{2/3}$.

Both NCS and NBW are implemented in {\small{OSIRIS}} with a Monte Carlo method. For NCS, at each time step photons are created randomly according to the total rate, and the corresponding energies are sampled according to the differential probability rate. The pair creation follows the same method. The momentum conservation is ensured by subtracting the momentum of the created photon from the lepton.
For NBW, the photon is removed from the simulation and its energy is distributed between the new electron and positron. 

For self-consistently simulating the quantum-dominated plasmas (or beams), the simulation time step ($\Delta t$) should be chosen such that to resolve both the collective plasma/beam physics and the SF-QED time scales. For example, to carefully model NCS, one has to satisfy $\Delta t\ll \tau_\gamma = \left(dP_\gamma/dt\right)^{-1}$.

\begin{figure*}
\includegraphics[width=14cm,height=8cm]{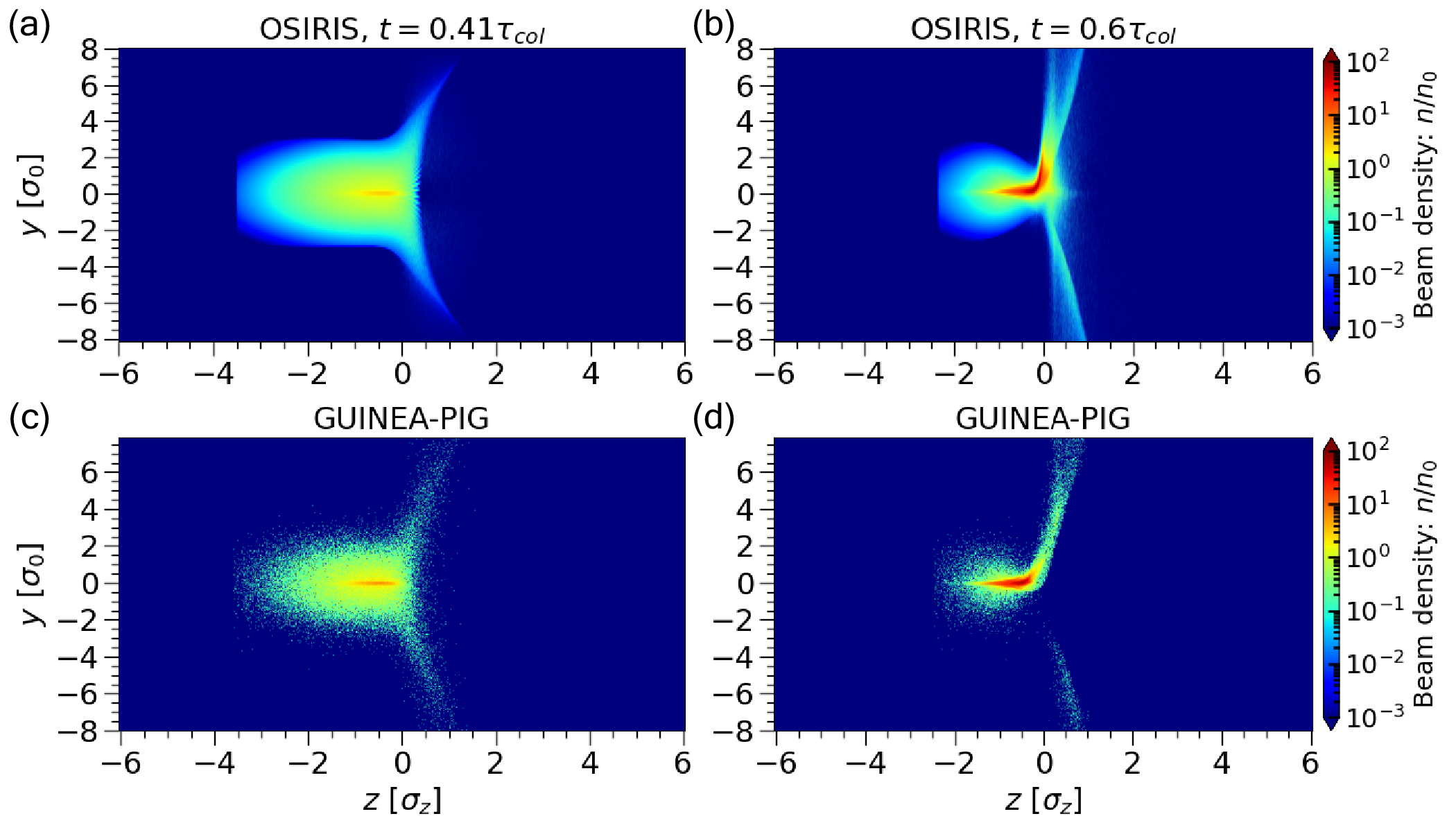}
\caption{(Color online). Simulations for a collision between Gaussian-profile electron beams, using codes of {\small{OSIRIS}} (upper row) and {\small{GUINEA-PIG}} (lower row), respectively. The beam parameters are given in Appendix \ref{SM_subsec: OS_GP}. The densities shown here are extracted from the $(y,z)$ slice across the beam center. (a) and (c) are taken shortly after the AP develops, where the beam pinch is already notable around the axis.}
\label{fig: OS_GP}
\end{figure*}

\section{Additional material for the discussions}
\label{SM_sec: additional_material_discussions}

\subsection{Profile transform between Gaussian beams and uniform, cylinder-shaped beams}
\label{SM_subsec: profile_transform}

A collision between Gaussian beams can be well approximated by an equivalent collision between uniform and cylinder-shaped beams \cite{Zhang2023}. The profile transform requires a stretch of beam length ($\sigma_z$), the conservation of SF-QED strength ($\chi_{e \, \mathrm{max}}$), and the peak particle flux (or current) between the Gaussian beams and the equivalent uniform beams. The particle flux is given by $\Gamma = 2\pi c\int nrdr$.

With the above requirements, the profile transform is given by
\begin{equation}
    \sigma_{z}^{\mathrm{U}}=2\sqrt{2}\sigma_{z}^{\mathrm{G}}, \ n_{0}^{\mathrm{U}}=0.41n_{0}^{\mathrm{G}}, \ \sigma_{0}^{\mathrm{U}}=2.22\sigma_{0}^{\mathrm{G}},
    \label{Eq: Suppl_profile_transform}
\end{equation}
where the superscript ``G" represents the Gaussian beam, and ``U" the equivalent uniform beam. Remarkably, this transform [Eq. \eqref{Eq: Suppl_profile_transform}] can also conserve the geometric luminosity ($L_0^{Geo}$). In addition, the corresponding disruption parameters satisfy
\begin{equation}
    D^{\mathrm{U}}=2.6D^{\mathrm{G}}.
    \label{Eq: Suppl_D_profile_transform}
\end{equation}
Equation \eqref{Eq: Suppl_profile_transform} and Eq. \eqref{Eq: Suppl_D_profile_transform} are useful to transform Eqs. \eqref{Eq: t_F}, \eqref{Eq: AP_condition_engineering}, and \eqref{Eq: theoretical_HD}) derived in this paper.

\begin{table*}[t] 
%\centering
\caption{Summary of the most prominent and feasible near-future projects of linear colliders where the designated parameters can allow for observing the anomalous pinch reported in this paper, i.e., the pinch condition (Eq. \eqref{Eq: AP_condition_engineering}) can be satisfied. Although the colliders here are primarily designed for $e^-e^+$ collisions, the $e^-e^-$ configuration is being considered as an alternative for the ultra-high-energy ($\sqrt{s}\ge \mathrm{TeV}$) particle physics study to rule out the technical difficulties with positron generation and acceleration \cite{Foster2025, ALEGRO2024, 10TeV_design_initiative}. $\sqrt{s}$ is the center-of-mass (c.o.m.) energy.}
\label{table: near_future_colliders}

\begin{tabular}[c]{|m{2.4cm}|m{3.4cm}|m{4.1cm}|m{3.7cm}|m{1.7cm}|}
	\hline
 
         \centering {{\textbf{Collider type}}} & \centering {{\textbf{Designated beam parameters}}} & \centering {{\textbf{Technology highlights}}} & \centering {{\textbf{Current status}}} & {{\textbf{Timeline}}} \\[9pt] 
        \hhline{|=|=|=|=|=|}

        \centering CLIC-based (CERN) \cite{Schulte2017, Shiltsev2021} \\ $ $ \\$ $ & \centering $\sqrt{s}=0.38\sim 3\ \mathrm{TeV}$\\ $\sigma_x=40\sim 150\ \mathrm{nm}$\\ $\sigma_y=1\sim 3\ \mathrm{nm}$\\ $\sigma_z=40\sim 70 \ \mathrm{\mu m}$\\ $N_0=(3\sim 6)\times 10^{9}$\\ $D=0.5\sim 150$\\ $\chi_e = 0.3\sim 23$  & \raggedright 1. Traditional radio-frequency (RF) based acceleration \\ 2. Novel two-beam acceleration scheme \\ 3. High accelerating field: $70\sim 100\ \mathrm{MV/m}$ & \raggedright 1. Conceptual Design Report delivered \cite{CLIC_CDR} \\ 2. Mature design and test study; Project approval $\sim 2028$; Tunnel construction $\sim 2030$ \cite{Stapnes2021} & $\sim 2040$ $~$ $~$ $~$ $~$ $~$ $~$ $~$ $~$ $~$ $~$ $~$ $~$ $~$ $~$ $~$ $~$ $~$ $~$ $~$ $~$  \\ [9pt]
	\hline  
        
        \centering $\mathrm{TeV}$-class advanced colliders \cite{Roser2023, Schroeder2023, Barklow2023} \\ $ $  & \centering $\sqrt{s}=1\sim 3\ \mathrm{TeV}$\\ $\sigma_x=10\sim 60\ \mathrm{nm}$\\ $\sigma_y\sim 1\ \mathrm{nm}$\\ $\sigma_z=5\sim 40\  \mathrm{\mu m}$\\ $N_0=(1\sim 5)\times 10^{9}$\\ $D=0.2\sim 100$\\ $\chi_e = 1\sim 600$  & \raggedright 1. (Staging) Plasma/dielectric-based wakefield acceleration \\ 2. Ultra-high accelerating field: $> \mathrm{GV/m}$ & \raggedright 1. Pre-project R$\&$D \\ 2. End-to-end design report $\sim 2028$ \\ $ $ \\ $ $ &  $2050$ $\sim$ $2070$ $~$ $~$ $~$ $~$ $~$ $~$ $~$ $~$ $~$ $~$ $~$ $~$ $~$ $~$ $~$ $~$ $~$ $~$ $~$ $~$ \\ [9pt]
	\hline
	
        \centering $10'\mathrm{s}\ \mathrm{TeV}$-class advanced colliders \cite{Roser2023, Schroeder2023, Barklow2023, 10TeV_design_initiative} \\ $ $ & \centering $\sqrt{s}=10\sim 30\ \mathrm{TeV}$\\ $\sigma_x=\sigma_y\sim 1\ \mathrm{nm}$\\ $\sigma_z=2\sim 40\ \mathrm{\mu m}$\\ $N_0=(3\sim 8)\times 10^{9}$\\ $D=0.6\sim 100$\\ $\chi_e = 800\sim 100000$  & \raggedright Round-beam scheme increases the luminosity-per-beam-power \cite{Roser2023, Barklow2023, Geddes2022} \\ $ $ & \raggedright 1. Recommended by the “P5 report” \cite{P5_Report} \\ 2. Pre-project R$\&$D \cite{10TeV_design_initiative} \\ 3. End-to-end design report $\sim 2028$ &  $2050$ $\sim$ $2070$ $~$ $~$ $~$ $~$ $~$ $~$ $~$ $~$ $~$ $~$ $~$ $~$ $~$ $~$ $~$ $~$ $~$ $~$ $~$ $~$ \\ [9pt]
	\hline
        
\end{tabular}

\end{table*}

\subsection{Comparison with {\small{GUINEA-PIG}}}
\label{SM_subsec: OS_GP}

{\small{GUINEA-PIG}} is a widely used code which is specialized for studying beam-beam collisions \cite{Schulte1996}. We have performed {\small{GUINEA-PIG}} simulations to benchmark our {\small{OSIRIS}} results. Here, we present a case study for a Gaussian-profile beam collision. The beams are cold with $\mathcal{E}_0=70 \ \mathrm{GeV}$, $N_0=4.05\times 10^9$, $\sigma_0=5.1\ \mathrm{nm}$, and $\sigma_z=6.5\ \mathrm{\mu m}$.

In the {\small{OSIRIS}} simulation, the numerical box is $23.7\sigma_0 \times 23.7\sigma_0 \times 12\sigma_{z}$ resolved by $200\times 200\times 72960$ grids, leading to the resolution of $\Delta x=\Delta y=0.11\sigma_0$ and $\Delta z=1.6\times 10^{-4}\sigma_z$. $1$ particle-per-cell (PPC) is used, corresponding to $7.3\times 10^{7}$ macro-particles per beam. The time step is $\Delta t=6\times 10^{-5}\sigma_z/c$. The beams are cut out for $r>3\sigma_0$ in radial direction, and for $|z-z_0|>3\sigma_z$ in $z$ direction, where $z_0$ is the beam center.

In the {\small{GUINEA-PIG}} simulation, the number of cells in transverse direction is $n_x=n_y=610$. The size of the simulation box is chosen such that the transverse grid resolution is the same with that of the {\small{OSIRIS}} simulation. The beams are divided into $128$ slices in longitudinal direction. The number of macro-particles is $2.5\times 10^6$ per beam. $10$ timesteps are used to move one slice of a beam to the next slice of the other beam.

When SF-QED is turned off, {\small{GUINEA-PIG}} gives the luminosity of $L_0=0.22L_0^{Geo}$, same as that computed in our {\small{OSIRIS}} simulation, where $L_0^{Geo}=N_0^2/4\pi \sigma_0^2$ is the geometric luminosity for Gaussian-beam collisions. When SF-QED is turned on, the anomalous pinch (AP) is observed in both simulations, as shown in Fig. \ref{fig: OS_GP}. In addition, the beams show similar dynamics. The deflection of the beams after they are severely pinched indicates the development of hosing/kink-like instabilities.

The time corresponding to the onset of the pinch $t_{\textrm{\tiny AP}}$ is measured to be the same in both codes. The luminosity recorded in {\small{GUINEA-PIG}} is slightly lower than that in {\small{OSIRIS}} which gives $L_0=0.46L_0^{Geo}$. In addition, the beam compression, the enhancement of the quantum parameter $\chi_e$, and pair production in {\small{GUINEA-PIG}} are also slightly lower than those in {\small{OSIRIS}}.

Therefore, our simulations have confirmed a reasonable agreement between {\small{GUINEA-PIG}} and {\small{OSIRIS}} for the $e^-e^-$ collision studied here, although with minor quantitative differences for the collision dynamics. We believe that these differences are due to the distinct simulation methodologies utilized in the two codes. However, this topic is beyond the scope of this paper. A detailed, systematic comparison between the {\small{GUINEA-PIG}} and {\small{OSIRIS}} simulations will be provided in a future publication.

As noted in the main text, the beams also gain a significant energy spread due to the beam-beam effects. Specifically, around the moments when AP is severe ($t\sim 0.5\tau_{col}$), the OSIRIS simulation shows that $\sim 15\%$ of the particles are within $1\%$ of the initial beam energy $\mathcal{E}_0$, and up to be $24\%$ of the particles within $10\%$ of $\mathcal{E}_0$. Therefore, there will be a trade-off between the overall luminosity enhancement and the broadened energy spread which can diminish the luminosity at the center-of-mass energy designated by the particle colliders. This can be studied in detail in future work.

\subsection{Impact of finite beam emittance}
\label{SM_subsec: impact_emittance}

The simulation results shown in the main text are obtained with perfectly collimated beams corresponding to zero emittance, to compare directly with our theory. We have also conducted the same QED-PIC simulations with finite-emittance beams (momentum divergence angle $\theta \sim \mathrm{mrad}$ or normalized emittance $\epsilon_n \sim \mathrm{mm}~\mathrm{mrad}$) without significant differences compared with the zero-emittance results ($\lesssim 10 \%$ difference). The zero-emittance predictions remain valid when the thermal divergence is smaller than the disruption-induced deflection \cite{Fabrizio2019, Zhang2023}, i.e., $\theta < D \sigma_0/\sigma_z$. This is verified for the regime investigated here with $D\gg 1$ and typical beam dimension of $\sigma_z\sim 10~\mathrm{\mu m}$ and $\sigma_0\gtrsim 10~\mathrm{nm}$.

\subsection{Near-future linear colliders suited for observing the anomalous pinch}
\label{SM_subsec: future_colliders}

The physical regime of AP discovered here can be reached by major next-generation $\mathrm{TeV}$-class lepton colliders. These colliders are either already scheduled or under extensive research and development (R$\&$D) in global collaborations \cite{Schulte2017, Shiltsev2021, Roser2023, Schroeder2023, Barklow2023, Geddes2022, P5_Report, Shiltsev2024, CLIC_CDR, Stapnes2021,10TeV_design_initiative}. We summarize in Table \ref{table: near_future_colliders} the near-future linear colliders where AP in $e^-e^-$ collisions will be present and must be carefully considered for the collider designs.

%\bibliographystyle{}
%\bibliography{refs}


\end{document}